\documentclass[12pt]{spieman}  
\usepackage{amsmath,amsfonts,amssymb}
\usepackage{graphicx}
\usepackage{setspace}
\usepackage[subfigure]{tocloft}
\usepackage{subfigure}
\usepackage{dsfont}
\usepackage{amssymb}
\usepackage{color}

\title{Identification and adaptive control of a high-contrast focal plane wavefront correction system}

\author[a, *]{He Sun}
\author[a]{N.Jeremy Kasdin}
\author[a, b]{Robert Vanderbei}
\affil[a]{Department of Mechanical and Aerospace Engineering, Princeton University, NJ, the United States.}
\affil[b]{Department of Operation Research and Financial Engineering, Princeton University, NJ, the United States.}

\cftpagenumbersoff{figure}
\cftpagenumbersoff{table} 
\begin{document} 
\maketitle

\begin{abstract}
All coronagraphic instruments for exoplanet high-contrast imaging need wavefront correction systems to reject optical aberrations and create sufficiently dark holes. Since the most efficient wavefront correction algorithms (controllers and estimators) are usually model-based, the modeling accuracy of the system influences the ultimate wavefront correction performance. Currently,  wavefront correction systems are typically approximated as linear systems using Fourier optics. However, the Fourier optics model is usually biased due to  inaccuracies in the layout measurements, the imperfect diagnoses of inherent optical aberrations, and a lack of knowledge of the deformable mirrors (actuator gains and influence functions). Moreover, the telescope optical system varies over time because of  instrument instabilities and  environmental effects. In this paper, we present an expectation-maximization (E-M) approach for identifying and real-time adapting the linear telescope model from data. By iterating between the E-step (a Kalman filter and a Rauch smoother) and the M-step (analytical or gradient-based optimization), the algorithm is able to recover the system even if the model depends on the electric fields, which are unmeasurable hidden variables. Simulations and experiments in Princeton's High Contrast Imaging Lab demonstrate that this algorithm improves the model accuracy and increases the efficiency and speed of the wavefront correction.
\end{abstract}


{\noindent \footnotesize\textcolor{black}{*}He Sun,  \linkable{hesun@princeton.edu} }

\begin{spacing}{1}   

\section{Introduction}
In the upcoming era of 30-meter ground-based telescopes and new advanced space telescopes, direct imaging is believed to be the next frontier in exoplanet detection and characterization. Unlike  indirect detection methods, such as radial velocity and transit, direct imaging collects light from the planet itself rather than its host star, thus enabling the spectral characterization of the planet's atmosphere and the full determination of its orbital parameters. But exoplanets are much fainter than their parent stars, requiring the \textcolor{black}{starlight's point spread function (PSF)} to be managed to make high contrast imaging of the exoplanet possible.

A leading technology for achieving the high contrast needed for exoplanet imaging is a coronagraph \cite{kasdin2003extrasolar, trauger2011hybrid, guyon2003phase, zimmerman2016shaped}. Consisting of a series of optimally designed masks and stops, coronagraphs are able to suppress the spread of starlight and thus create high-contrast detection regions, so-called dark holes, in the image plane. However, since the coronagraphs are designed assuming perfect optics, they are fundamentally sensitive to any wavefront perturbations. \textcolor{black}{Even small aberrations introduce bright stellar speckles in the dark holes, which influence the instrument ability for exoplanet obervations.} To maintain a high contrast for exoplanet observations, wavefront correction is required for all coronagraph instruments. In a ground-based telescope, the wavefront correction system typically includes a wavefront sensor, such as a spatially filtered Shack-Hartmann sensor or a pyramid sensor\cite{verinaud2005adaptive, frazin2017fast}, in the pupil plane to measure the wavefront aberrations and then directly compensates for them using deformable mirrors (DMs). Such a system is able to cancel pupil phase aberrations due to atmospheric turbulence and achieve contrasts of $10^{-5}$ to $10^{-6}$ on current telescopes, allowing for the imaging of young hot gas giant planets\cite{macintosh2014first}. 
Directly imaging dimmer planets (down to Earth size) at higher contrast requires a space telescope that reaches contrasts below $10^{-8}$ before post-processing\cite{spergel2015wide, noecker2016coronagraph}.  For these coronagraph instruments targeting earth-sized planets, the pupil plane approach with a separate wavefront sensor is not capable of reaching the required high contrast values because of non-common-path errors. Instead, we need to estimate the focal-plane electric field using only camera images and compute the DM control signals for based on the estimated field. This estimation and control problem is commonly referred to as focal-plane wavefront correction (FPWC). Effective FPWC algorithms require efficient estimation algorithms and accurate models of the optical system, particularly of the influence of DM voltage commands on the focal-plane electric field.

In all current FPWC systems, the optical models needed for control and estimation have been derived by applying Fourier optics to the optical layout. However, using only the Fourier optics approach results in significant bias errors due to inaccuracies in measurements of the optical system,  imperfect knowledge of the systematic optical aberrations, and poor or biased models of the DM influence.  This has a detrimental effect on  the wavefront correction speed and the final achievable contrast. \textcolor{black}{Classical approaches for eliminating these model errors and improving the system performance include pupil plane phase retrieval\cite{fienup1982phase, shechtman2015phase} and laser interferometric DM surface characterization\cite{prada2017characterization} in advance, which are usually time consuming and also introduce non-common-path errors.} In this paper, we propose a new data-driven framework using the expectation-maximization (E-M) algorithm to accurately identify and adaptively control the FPWC system. In contrast to classical approaches, our method does not require a change to the optical system design, tracks real-time systematic changes, and speeds up convergence of the controller.

In the following sections, we first provide a brief overview of the FPWC system, including mathematical modeling and current state of the art on wavefront estimation, control, and \textcolor{black}{model calibration}. In addition, we also propose a new idea to formulate the problem as a stochastic optimization problem. Then we review the E-M algorithm and derive the E-M equations of the FPWC system. We finally report the simulation and experimental results on the FPWC system identification and adaptive control in the Princeton High Contrast Imaging Lab (HCIL) to demonstrate the method's ability. 

\section{Overview of high-contrast focal plane wavefront correction}
\textcolor{black}{In this section, we review the current state of the art in FPWC and we also introduce a new idea to formulate FPWC as a stochastic optimization problem. In Sec.~\ref{subsec:model}, \ref{subsec:estimationControl} and \ref{subsec:systemID}, we review the approaches of optical system modeling, wavefront estimation and control, and model calibration, which are related to our new algorithm. Readers already familiar with these subjects may skip these sections.} 
\label{sec:overview}
\subsection{Mathematical modeling} \label{subsec:model}
The FPWC system with a coronagraph is typically formulated as a state-space model for the convenience of control applications. In this state-space formulation, the control inputs, observations and state variables are respectively the DM voltage commands, camera images, and the focal plane electric fields. We begin this section by deriving this underlying state-space model.

\begin{figure}[!htb]
	\centering
	\includegraphics[width=6in]{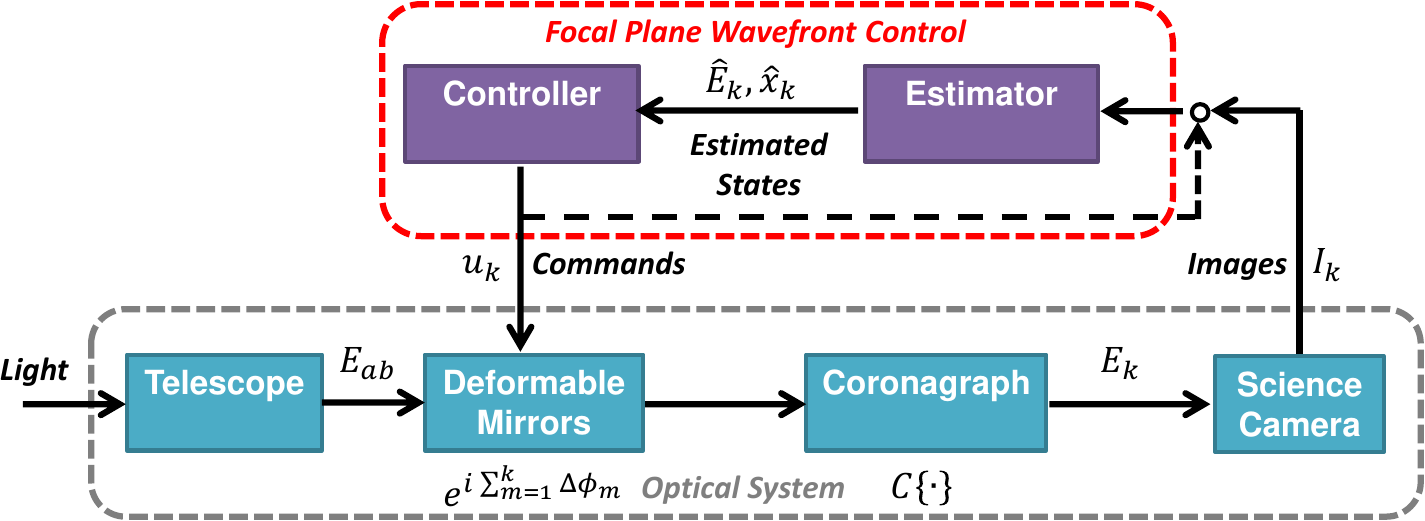}
\caption{Telescope optical system architecture and focal plane wavefront control loop. System variables are also marked on the diagram, where $E_{ab}$ is the aberrated pupil plane electric field, $\Delta \phi_m$ is the DM surface phase change at a single step, $\mathcal{C}\{\cdot\}$ represents the light propagation through coronagraph, $E_k$ is the focal plane electric field, $I_k$ represents the camera images, $u_k$ represents the DM commands, and $\hat{E}_k$/$\hat{x}_k$ are the estimated complex/real-valued states of electric field.} 
\label{fig:controlSystem}
\end{figure}

The block diagram in Fig.~\ref{fig:controlSystem} shows the  architecture of the telescope optics and the control loop. We define \textcolor{black}{$E_{ab}$ as the aberrated pupil electric field} before the DMs and $\Delta \phi_m$ as the phase change introduced by the DMs at correction iteration $m$. The coronagraph operator $\mathcal{C}\{\cdot\}$ represents the propagation from the DM to the focal plane camera. After $k$ correction iterations, the focal plane electric field is given by,
\textcolor{black}{
\begin{equation}
E_k = \mathcal{C}\{E_{ab} \exp(i \sum_{m=1}^k \Delta \phi_m)\}.
\label{eq:propagation}
\end{equation}}
When the phase change due to the DM is small (typically the DM surface perturbation is smaller than 30nm), the focal plane electric field in Eq.~\ref{eq:propagation} can be expanded in a Taylor series about $\Delta \phi_m$ to yield
\begin{equation} \label{eq:Taylor}
\begin{aligned}
E_k &\approx \mathcal{C}\{E_{ab}\} + \sum_{m=1}^k \mathcal{C}\{E_{ab} i \Delta \phi_m\}\\
& = E_{k-1} + \mathcal{C}\{E_{ab} i \Delta \phi_k\}.
\end{aligned}
\end{equation}
where we used the fact that the coronagraph is a linear operator in the applied electric field (as it is composed of Fresnel propagations, Fourier transforms, and coronagraph mask multiplications). 

The DM phase change, $\Delta \phi_k$, at each step is approximated by summing weighted influence functions produced by each actuator on the DM. That is, given an influence function, $ f_q$, which represents the $q$-th actuator's response to a unit voltage input, the DM induced phase change across the pupil is approximated by the linear superposition
\begin{equation} \label{eq:influenceFun}
\Delta \phi_k \approx \sum_{q=1}^{N_{act}} u_{k,q} f_q,
\end{equation}
where $N_{act}$ is the number of DM actuators and $u_{k, q}$ is the voltage command change of the q-th actuator. \textcolor{black}{We use $u_{k, q}$ instead of $\Delta u_{k, q}$ in this paper for notational simplicity.}

Substituting Eq.~\ref{eq:influenceFun} into Eq.~\ref{eq:Taylor} results in a linear relationship between the focal plane electric field and the DM voltage commands,
\begin{equation} \label{eq:linearRelation}
E_k = E_{k-1} + \sum_{q=1}^{N_{act}} u_{k,q} \mathcal{C}\{E_{ab} i f_q\}.
\end{equation}

By discretizing and vectorizing the 2-D electric fields, Eq.~\ref{eq:linearRelation} can be put in the common matrix form of the state transition model,
\begin{equation} \label{eq:stateTransition}
E_k = E_{k-1} + F(E_{ab}, f_{1:N_{act}}) u_k,
\end{equation}
where $E_{k}, E_{k-1} \in \mathbb{C}^{N_{pix} \times 1}$ are the complex state vectors, $u_k \in \mathbb{R}^{N_{act} \times 1}$ is the DM control input vector, $F \in \mathbb{C}^{N_{pix} \times N_{act}}$ is the system Jacobian matrix, and $N_{pix}$ is the number of camera pixels in the dark holes. The corresponding intensity on the science camera is,
\begin{equation} \label{eq:intensity}
I_k = E_k^{\star} \circ E_k,
\end{equation}
where $\circ$ represents the element-wise multiplication and $\star$ denotes complex conjugation. Eq.~\ref{eq:stateTransition} and Eq.~\ref{eq:intensity} are, respectively, the state transition and observation equation of the state-space model.

Since the measurement Eq.~\ref{eq:intensity} is the sum of the squares of the real and imaginary components, it is impossible to extract the full complex electric field from a single measurement.  Instead, the current approach is to apply $n$ ($n \geq 2$) \textcolor{black}{pairs of opposite} ``probe'' commands to the DM\cite{borde2006high, give2011pair}, denoted by $u_k^p = [u_k^{p, 1}, \cdots, u_k^{p, n}]$, which result in the set of $2n$ intensity measurements,  
%
%
\begin{equation} 
\begin{aligned}
&I_k^{m+} =  (E_{k} + F u_k^{p, m})^{\star} \circ (E_{k} + F u_k^{p, m}), \quad \forall m = 1, \cdots, n\\
&I_k^{m-} =  (E_{k} - F u_k^{p, m})^{\star} \circ (E_{k} - F u_k^{p, m}), \quad \forall m = 1, \cdots, n.\\
\end{aligned}
\end{equation}
These are then subtracted to form an overdetermined set of $n$ linear measurements of the electric field,
\begin{equation} \label{eq:differenceImage}
\begin{bmatrix}
\Delta I_k^1\\
\vdots\\
\Delta I_k^n
\end{bmatrix}
=
\begin{bmatrix}
I_k^{1+} - I_k^{1-}\\
\vdots\\
I_k^{n+} - I_k^{n-}
\end{bmatrix}
= \Re\left\{
\begin{bmatrix}
4 (F u_k^{p, 1})^{\star} \circ E_k\\
\vdots\\
4 (F u_k^{p, n})^{\star} \circ E_k
\end{bmatrix}\right\}
= \Re\left\{
\begin{bmatrix}
4 \text{diag}\{(F u_k^{p, 1})^{\star}\}\\
\vdots\\
4 \text{diag}\{(F u_k^{p, n})^{\star}\}
\end{bmatrix} E_k \right\}, 
\end{equation}
where $\text{diag}\{\cdot\}$ represents the diagonal matrix constructed from a vector. Equations~\ref{eq:stateTransition} and \ref{eq:differenceImage} make up the linear state-space model of the FPWC system.

The element-wise product structure in Eq.~\ref{eq:differenceImage} decouples the linear state transition equations in each pixel, so the electric field of a single pixel can be estimated  based only on its own intensity measurements. For mathematical convenience, we can further split the real and imaginary part of the electric field and derive real-valued state-space equations,
\begin{equation} \label{eq:stateSpace}
\begin{aligned}
x_{k, j} &= x_{k-1, j} + G_j u_k,\\
z_{k, j} & = H_{k, j} x_{k,j},
\end{aligned}
\end{equation}
where $j \in \{1, \cdots, N_{pix}\}$ is the index of camera pixels, and
\begin{equation} \label{eq:realState}
x_{k, j} = 
\begin{bmatrix}
\Re\{E_{k, j}\}\\
\Im\{E_{k, j}\}
\end{bmatrix},
G_j = 
\begin{bmatrix}
\Re\{F_{j, 1:N_{act}}\}\\
\Im\{F_{j, 1:N_{act}}\}
\end{bmatrix},
z_{k, j} =
\begin{bmatrix}
\Delta I_{k, j}^1\\
\vdots\\
\Delta I_{k, j}^n
\end{bmatrix},
H_{k, j} = 4 (G_j u_k^p)^T.
\end{equation}
The elements of the state vectors and matrices are now real numbers, which is more convenient for developing estimators and controllers.

\textcolor{black}{Good DM ``probe'' commands should help construct well-conditioned measurement matrices, $H_{k, j}$, for all the pixels in the dark holes. Commands that create ``Sinc'' waves on the DM surface are usually good choices, because camera pixels in two symmetric rectangular areas are influenced by this type of probe according to Fourier analysis.} 

\subsection{Focal plane wavefront estimation and control} \label{subsec:estimationControl}
With the state-space model developed in Sec.~\ref{subsec:model}, we now introduce the wavefront estimation and control algorithms.
The baseline wavefront estimation approach used in most implementations to date is the least-square, batch process estimator (BPE)\cite{borde2006high, give2011pair}, which can be derived as follows. We begin with the linear observation model in Eq.~\ref{eq:stateSpace} but with an additive noise term,  $n_{k,j}$, to represent
camera measurement noise and observation matrix errors (originally from Jacobian matrix errors),
\begin{equation} \label{eq:observationNoise}
z_{k, j} = H_{k, j} x_{k, j} + n_{k, j}.
\end{equation}
\textcolor{black}{Theoretically, the camera measurements should follow Possion distributions. However, in our case there are a sufficient number of starlight photons for detection, making it safe to assume the measurements follow centered Gaussian distributions on the top of the speckles, i.e. $n_{k, j} \sim \mathcal{N}(0, R_{k, j})$.} We can thus perform  a least-square regression to estimate the expectation, $\hat{x}_{k, j}$, and the covariance matrix, $P_{k, j}$, of the state vector at each time step, $k$,
\begin{equation} \label{eq:lse}
\begin{aligned}
\hat{x}_{k, j} &= (H_{k, j}^{T} H_{k, j})^{-1} H_{k, j}^T z_{k, j},\\
P_{k, j} &= (H_{k, j}^T H_{k, j})^{-1} H_{k, j}^T R_{k, j} H_{k, j} (H_{k, j}^T H_{k, j})^{-1}.
\end{aligned}
\end{equation}
Repeating this regression procedure for all pixels provides an estimate of the entire electric field in the focal plane. \textcolor{black}{Although this algorithm is easy to implement, one weakness is that its estimation accuracy, indicated by the estimation covariance, is fully determined by the measurement noises.} When the signal-noise-ratio (SNR) is low (which happens as the dark hole improves), this batch process estimator may not provide accurate enough estimates to be used for control.\textcolor{black}{\cite{riggs2016recursive}}

A better solution is to incorporate  prior knowledge from the model and the previous measurements using a Kalman filter\cite{groff2013kalman}. This formulation allows us to introduce an additive process noise term, $w_k$, to the state-space equations,
\begin{equation} \label{eq:transitionNoise}
x_{k, j} = x_{k-1, j} + G_j u_k + w_{k, j},
\end{equation}
\textcolor{black}{where $w_{k, j} \cong \Delta G_j u_k + r_{k, j}$, of which the first term comes from the Jacobian matrix errors and the second term comes from system instabilities, such as DM drift. Assuming elements of the Jacobian matrix bias $\Delta G_j$ and the instability term $r_{k, j}$ all follow zero-mean Gaussian distributions\footnote{\textcolor{black}{The model bias $\Delta G_j$ is the difference between the true Jacobian matrix and the current Jacobian matrix in use. Each element of Jacobian errors, $\Delta G_j$, has zero mean, since we have no knowledge whether the current Jacobian influence is larger or smaller than the true value. This is also intuitively true in real experiment because the biases of different actuators may have influence of different directions, so they will cancel with each other.}}, the process noise also becomes an additive Gaussian noise, $w_{k, j} \sim \mathcal{N}(0, Q_{k, j})$, which satisfies the requirement of Kalman filtering.}

At  control iteration $k$, the state transition model provides a prediction of the current state, since we have $\hat{x}_{k-1, j}$ and $P_{k-1, j}$ from the previous estimation. With this prior knowledge, we can derive the log-likelihood function of the current state and the observations,
\begin{equation} \label{eq:loglikelihood}
\begin{aligned}
\log \ p(z_{k, j}, x_{k, j}) = &-\frac{1}{2} (x_{k, j} - \hat{x}_{k, j|k-1})^T P_{k, j|k-1}^{-1} (x_{k, j} - \hat{x}_{k, j|k-1})\\
&- \frac{1}{2} (z_{k, j} - H_{k, j} x_{k, j})^T R_{k, j}^{-1} (z_{k, j} - H_{k, j} x_{k, j}),
\end{aligned}
\end{equation}
where
\begin{equation}
\begin{aligned}
\hat{x}_{k, j|k-1} &= \hat{x}_{k-1, j} + G_j u_k,\\
P_{k, j|k-1} &= H_{k, j} P_{k, j} H_{k, j}^T + R_{k, j}
\end{aligned}
\end{equation}
are respectively the a-priori state and covariance estimates and $p(z,x)$ is the joint probability density function for $z$ and $x$.
The Kalman filter maximizes this log-likelihood function, so it optimally combines the information from the model and the observations and thus reduces the estimation covariance.

Recently, Riggs et al.\cite{riggs2016recursive} introduced
an incoherent light term into the nonlinear observation model, Eq.~\ref{eq:intensity},
\begin{equation} \label{eq:incohrent}
I_k = E_k^{\star} \circ E_k + I_{inco, k}.
\end{equation}
They then employed an extended Kalman filter (EKF) to simultaneously estimate both the coherent electric field and incoherent intensity (which contains the planet). This method removes the requirement for pair-wise probing and makes simultaneous wavefront correction and planet detection possible. This EKF approach is not used in this paper, however, it will be applied to improve the system identification work described here in a future paper to characterize system nonlinearities.
 
With the state estimates available, it is now possible to compute the DM voltage commands to manipulate the focal plane electric field. For mathematical simplicity, we construct a real, linear state transition model by combining the state equations for each pixel (Eq.~\ref{eq:transitionNoise}) into a single vectorized form,
\begin{equation}  \label{eq:transitionReal}
x_{k} = x_{k-1} + G u_k + w_{k},
\end{equation}
where
\begin{equation} \label{eq:realState2}
x_{k} =
\begin{bmatrix}
x_{k, 1}\\
\vdots\\
x_{k, N_{pix}}
\end{bmatrix},
G =
\begin{bmatrix}
G_{1}\\
\vdots\\
G_{N_{pix}}
\end{bmatrix},
w_{k} =
\begin{bmatrix}
w_{k, 1}\\
\vdots\\
w_{k, N_{pix}}
\end{bmatrix}.
\end{equation}
The controller must drive the state $x_k$ as close to zero as possible to maintain a high contrast in the dark hole. Currently, the two most popular model-based optimal controllers are electric field conjugation (EFC)\cite{give2007broadband} and stroke minimization (SM)\cite{pueyo2009optimal}.

EFC works by minimizing a cost function consisting of the total energy in the dark holes and a Tikhonov regularization, which can be written,
\begin{equation} \label{eq:efc}
\min_{u_k} \quad x_{k}^{T} x_{k} + \alpha_k u_k^T u_k, \quad s.t. \quad x_{k} = x_{k-1} + G u_k,
\end{equation}
where $\alpha_k$ is the Tikhonov regularization parameter.

In contrast, stroke minimization aims to find the smallest DM commands that achieve a target contrast, which can be formulated as the constrained minimization,
\begin{equation} \label{eq:sm1}
\min_{u_k} \quad u_k^T u_k, \quad s.t. \quad x_k^{T} x_k = C_k, \quad x_{k} = x_{k-1} + G u_k,
\end{equation}
where $C_k$ is the target contrast, or total energy, in the dark holes. The equality constraints can be incorporated into the cost function via a Lagrange multiplier, making stroke minimization into a similar formula to EFC,
\begin{equation} \label{eq:sm2}
\min_{u_k} \quad u_k^T u_k + \mu_k (x_k^{T} x_k - C_k), \quad s.t. \quad x_{k} = x_{k-1} + G u_k.
\end{equation}

The optimal solutions of Eq.~\ref{eq:efc} and Eq.~\ref{eq:sm2} give two corresponding feedback control laws,
\begin{equation} \label{eq:efcsm}
u_k = - (G^{T}G + \alpha_k \mathbb{I})^{-1} G^{T}x_{k-1} \quad \text{and} \quad u_k = - (G^{T}G + \frac{1}{\mu_k} \mathbb{I})^{-1} G^{T} x_{k-1}.
\end{equation} 
where $\mathbb{I} \in \mathbb{R}^{N_{act} \times N_{act}}$ is the identity matrix.

It is evident that EFC and stroke minimization, as more rigorously discussed by Groff et al.\cite{groff2016methods}, in fact define the same control law except for the tuning parameters, $\alpha_k$, and the Lagrange multiplier, $\mu_k$. The Lagrange multiplier, $\mu_k$, is a function of the target contrast $C_k$, which is the tuning parameter in stroke minimization.  Both $\alpha_k$ and $\mu_k$ introduce a damping term, although based on different considerations, in the matrix inversion, which helps avoid an ill-posed matrix inversion problem and, more importantly, reduce the influence of Jacobian matrix biases. Tuning the damping parameter, $\alpha_k$ and $\mu_k$, which turns out to be non-trivial, is the key to properly implementing the controllers.\cite{marx2017electric}

\subsection{Model calibration} \label{subsec:systemID}
\textcolor{black}{Because the wavefront estimators and controllers are all model-based, their performance highly depends on the accuracy of the underlying model. It is common to pre-calibrate the model based on some testbed measurments before running high-contrast focal plane wavefront correction. To date, all the model calibration approaches work to improve the Jacobian matrix in the linear state-space formulation.}

As indicated by Eq.~\ref{eq:stateTransition}, the Jacobian matrix is fundamentally a function of the aberrated pupil electric field, $E_{ab}$, and the actuator influence functions, $f_{1:N_{act}}$, so an indirect approach to improving the model is to characterize $E_{ab}$ and $f_{1:N_{act}}$ separately and then compute the Jacobian matrix based on the coronagraphic propagation equation \textcolor{black}{in Eq.~\ref{eq:linearRelation}}. The influence functions are usually characterized using laser interferometry\cite{prada2017characterization}. Since it is too time consuming to measure the surface responses of all the actuators (several thousands on each DM), typically only a few representative actuators are characterized with the assumption that all actuators have similar responses. The pupil electric field, though, cannot be directly measured.  It is typically reconstructed from multiple focused and defocused images using phase retrieval algorithms\textcolor{black}{\cite{marx2016phase, sauvage2012coronagraphic, jurling2014applications, paine2018machine}}. However, all the phase retrieval algorithms assume a certain light propagation model, so they do not have the ability to diagnose any errors from an incorrect optical layout prescription. In addition, since the coronagraph typically blocks most of the light from the entrance pupil, there are very few photons to provide the needed information.  \textcolor{black}{To fix this, current phase retrieval approaches require removal of the coronagraph to collect data; this makes the phase retrieval time consuming and prone to non-common path error.}

Recent work by Zhou et al. \textcolor{black}{started exploring system identification methods} for determining  the Jacobian matrix in favor of directly identifying the Jacobian matrix by perturbing the DM shapes and observing the resulting camera images\cite{zhou2016closing}. The physical interpretation of a Jacobian matrix column is the influence of a DM actuator with unit voltage command on the focal plane electric field. Therefore, by definition, the Jacobian matrix can be derived by commanding each actuator and estimating the focal plane electric field changes. The least-squared, batch process estimator (BPE) was employed in that work for the electric field estimation. However, since BPE requires a large amount of data and is relatively noisy, the identification procedure was time consuming and the resulting identified model was too noisy to be used in the wavefront correction. In addition, the identified model using BPE was also limited by the initial knowledge of the Jacobian matrix. Therefore, up to now, this work has only been used for qualitatively understanding the sources of the model errors, instead of quantitatively correcting the Jacobian matrix errors.

\subsection{New theoretical results: FPWC as a stochastic optimization problem} \label{sec:stochasticOpt}
As can been seen in Sec.~\ref{subsec:estimationControl} and \ref{subsec:systemID}, the typical approach to focal-plane wavefront control is to examine the wavefront estimation, wavefront control, and model calibration as separate problems. In this section, we try to bridge these aspects by formulating the FPWC problem as a single stochastic optimization problem. As first shown by Sun et al.\cite{sun2017improved}, this approach provides better physical insights into the tuning parameters in the algorithms and also provides theoretical ayalyses on how the wavefront control, estimation, and model accuracy influence the final contrast in the dark hole.

The ultimate goal of the FPWC is to minimize the total intensity, $x_k^T x_k$, in the dark holes. Since the state, $x_k$, is a random variable, we can formulate FPWC as a stochastic optimization/control problem  that minimizes the expectation of the dark hole intensity, $\left<x_k^{T} x_k\right>$. The state variable follows the stochastic process in Eq.~\ref{eq:transitionReal}. With the assumption that the process noise, $w_k$, has a zero mean, the expectation at step $k$ can be distributed as
\begin{equation} \label{eq:stochasticoptimization}
\begin{aligned}
\left<x_k^{T} x_k\right> &= \left<x_{k-1}^T x_{k-1}\right> + 2 \left<x_{k-1}\right>^{T} G u_k + u_k G^{T} G u_k + \left<w_k^{T} w_k\right>\\
 &= \left<x_{k-1}^T \right> \left<x_{k-1}\right> + 2 \left<x_{k-1}\right>^{T} G u_k + u_k G^{T} G u_k + \left<w_k^{T} w_k\right> + \sum_{j = 1}^{N_{pix}} \mbox{Tr}(\text{var}(x_{k-1, j})), \\
\end{aligned}
\end{equation}
where the statistics of the previous state are provided by the past wavefront estimation, $\left<x_{k-1} \right> = \hat{x}_{k-1}$ and $\text{var}(x_{k-1, j}) = P_{k-1, j}$.\footnote{\textcolor{black}{The covariance matrix $P_{k-1, j}$ is an indicator for the estimation accuracy, which is also a function of $\hat{x}_{k-1}$.}}

\textcolor{black}{The process noise, as explained in Sec.~\ref{subsec:estimationControl}, includes the Jacobian matrix errors and the system instabilities, $w_k \cong \Delta G u_k + r_k$. Given $r_k \sim \mathcal{N}(0, S_{k})$, the process noise covariance in Eq.~\ref{eq:stochasticoptimization} becomes $\left<w_k^{T} w_k\right> = u_k^T \left<\Delta G^T \Delta G \right> u_k + S_k$, where $\left<\Delta G^T \Delta G \right> \triangleq W$ models Jacobian uncertainties.}

The stochastic optimization problem can now be written
\textcolor{black}{
\begin{equation} \label{eq:stochasticoptimization2}
\min_{\hat{x}_{k-1}, u_k} \ \Phi(\hat{x}_{k-1}, u_k) = \hat{x}_{k-1}^T \hat{x}_{k-1} + 2 \hat{x}_{k-1}^T G u_k + u_k G^{T} G u_k +  u_k^T W u_k + \sum_{j = 1}^{N_{pix}} \mbox{Tr}(P_{k-1, j}).
\end{equation}}
\textcolor{black}{$S_k$ is eliminated in the optimization because it is a constant covariance matrix.} As this cost function indicates, the final contrast depends on not only the DM commands, $u_k$, but also the estimation accuracy, $\mbox{Tr}(P_{k-1, j})$, and the Jacobian uncertainties, $W$. Minimizing the first four terms of the cost function over $u_k$ defines the wavefront controller, while minimizing the trace of the estimation covariance matrix, $\mbox{Tr}(P_{k-1, j})$, over $\hat{x}_{k-1}$ defines the wavefront estimator. In addition, system
identification or classical model calibration can be used to reduce the model uncertainties, $\mbox{Tr}(W) = \|\Delta G\|_F^2$, which also improves the final achievable contrast from the wavefront correction.

By definition, the entries of the regularization matrix are
\begin{equation} \label{eq:MatrixRegularization}
W_{m, l} = \sum_{j=1}^{N_{pix}} \left< \Delta G_{j, m}^T \Delta G_{j, l} \right>, \quad \Delta G_j \in \mathbb{R}^{2 \times N_{act}}, \forall j = 1, \cdots, N_{pix},
\end{equation}
where the subscripts, $m$ and $l$, represent the column indices of the Jacobian bias matrices. Each column of $\Delta G$ gives the modeling errors of an actuator's influence, so $W_{m, l}$ indicates the covariance of Jacobian errors from the $m$-th and $l$-th actuators. In general, $W$ is a symmetric positive definite matrix with nearly all the entries non-zeros.

The off-diagonal entries in $W$ disappear if \textcolor{black}{the modeling errors of} different actuators are assumed to be \textcolor{black}{unrelated} from each other. EFC or SM with scalar regularization further assume that the covariance of errors from different actuators are identical. Thus, given that
\begin{equation} \label{eq:errstd}
\begin{aligned}
\left< \Delta G_{j, m}^T \Delta G_{j, m} \right> &= \mbox{Tr}(\text{var}(\Delta G_{j, m})) = 2 \sigma^2, \forall j, m, \\
\left< \Delta G_{j, m}^T \Delta G_{j, l} \right> &=\mbox{Tr}(\text{cov}(\Delta G_{j, m}, \Delta G_{j, l})) = 0, \forall j, m \neq l, 
\end{aligned}
\end{equation}
$W$ degrades to a scaled identify matrix,
\begin{equation} \label{eq:regularization}
W = 2 N_{pix} \sigma^2 \mathbb{I}.
\end{equation}
This shows that tuning the Tikhonov regularization parameter or Lagrange multiplier is equivalent to finding the magnitude of Jacobian uncertainties in our model. A smaller regularization parameter indicates smaller Jacobian errors, which finally leads to higher contrast according to Eq.~\ref{eq:stochasticoptimization2}. In the following sections, we will present the system identification and the adaptive control using the scalar regularization assumption in Eq.~\ref{eq:errstd}. \textcolor{black}{This assumption is not fundamentally necessary for our E-M algorithm, but it will significantly simplify the algorithm implementation.} Characterizing the filled regularization matrix \textcolor{black}{(by assuming each actuator's error not independent)} turns out to be very hard, because the high-dimensional system it suggests is usually underdetermined and requires tremendous amount of data for identification. To proceed with this idea, we have to assume some known structure of the regularization matrix\footnote{For example, we can assume only neighboring actuators are coupled, which makes the matrix very sparse.} \textcolor{black}{or incorporate dimension reduction techniques, such as principal component analysis (PCA) or sigular-value decomposition (SVD),} to reduce the number of adaptable parameters. We will leave these explorations for future work.

\section{Expectation-Maximization (E-M) algorithm} \label{sec:EM}
The new stochastic optimization formulation in Sec.~\ref{sec:stochasticOpt} indicates the potential from system identification for improving the wavefront corrections. Moreover, it indicates not only that identifying the Jacobian matrix is necessary, but also that characterizing the process  and observation noises are important for tuning the optimal estimators and controllers. In this section, we develop a new E-M algorithm based approach\cite{sun2017identification} to accomplish all of these goals. 

\subsection{Review of the E-M algorithm}
The E-M algorithm is an iterative system identification algorithm to find the maximum a posteriori (MAP) estimates of the model parameters in the presence of hidden variables\textcolor{black}{\cite{dempster1977maximum, murphy2014machine}}. \textcolor{black}{Hidden variables are the states of a dynamical system which are not directly observable.} Since the true values of the hidden variables are absent, we cannot explicitly derive the log-likelihood function and apply the maximum likelihood estimation (MLE) to identify the model parameters as usual. Instead, we maximize a lower bound of the log-likelihood of only the model inputs and outputs. In general, with the hidden variables, the model inputs and outputs (\textcolor{black}{the commands and observations of a system, usually referred to as the training data}), and the model parameters (\textcolor{black}{coefficients parametrizing the model function}), denoted as $X$, $Y$ and $\theta$ respectively, the log-likelihood can be written as an integral of the marginal probability over the hidden variables,
\begin{equation} \label{eq:likelihood}
\mathcal{L}\{\theta\} = \log \ p(Y|\theta) = \log \int p(X, Y | \theta) dX.
\end{equation}

Assuming \textcolor{black}{the hidden variables follow a probability distribution}, $\mathcal{Q}(X)$, a lower bound on the log-likelihood, $\mathcal{F}(\mathcal{Q}, \theta)$, can be found using Jensen's inequality,
\begin{subequations} \label{eq:lowerbound}
\begin{align}
\mathcal{L}\{\theta\} &= \log \ p(Y|\theta) = \log \int p(X, Y | \theta) dX\\
&= \log \int \mathcal{Q}(X) \ \frac{p(X, Y | \theta)}{\mathcal{Q}(X)} dX\\
&\geq \int \mathcal{Q}(X)  \ \log \ \frac{p(X, Y | \theta)}{\mathcal{Q}(X)} dX\\
&= \int \mathcal{Q}(X) \  \log \ p(X, Y | \theta) dX - \int \mathcal{Q}(X)  \log \ \mathcal{Q}(X) dX\\
&= \mathcal{F}(\mathcal{Q}, \theta).
\end{align}
\end{subequations}

The E-M algorithm alternates between maximizing this lower bound with respect to the hidden variable distribution, $\mathcal{Q}(X)$, and the model parameters, $\theta$. Optimizing over the distribution $\mathcal{Q}(X)$ while fixing  $\theta$ is called the expectation-step (E-step), and Optiming over the model parameters $\theta$ while fixing  $\mathcal{Q}(X)$ is called the maximization-step (M-step).

\textcolor{black}{In the E-step,  $\mathcal{F}(\mathcal{Q}, \theta)$ is maximized when the inequality in Eq.~\ref{eq:lowerbound}(c) becomes an equality, i.e. $\mathcal{L}\{\theta\} = \mathcal{F}(\mathcal{Q}, \theta)$. Equality in Eq.~\ref{eq:lowerbound}(c) holds if and only if $p(X, Y | \theta)/\mathcal{Q}(X)$ is constant for any possible $X$. The joint probability $p(X, Y | \theta)$ can be rewritten as a conditional probability using Bayes' rule,
\begin{equation} \label{eq:jointdistribution}
p(X, Y | \theta) = p(X| Y, \theta) p(Y| \theta),
\end{equation}
so $\mathcal{F}(\mathcal{Q}, \theta)$ is maximized when $\mathcal{Q}(X)=p(X| Y, \theta)$, since $p(Y| \theta)=p(X, Y | \theta)/p(X| Y, \theta)$ does not depend on $X$.}

\textcolor{black}{In the M-step, $\mathcal{F}(\mathcal{Q}, \theta)$ is maximized when $\int_X \mathcal{Q}(X) \  \log \ p(X, Y | \theta) dX = \mathrm{E}_X[\log \ p(X, Y | \theta)]$, the expectation of the log likelihood, is maximized. This is a stochastic maximum likelihood estimation (MLE) problem of the model parameters, $\theta$.}

\textcolor{black}{Theoretically, the model parameter estimation always converges to a local minimum after enough iterations of the E-step and the M-step. The number of iterations it takes depends on the initial knowledge of the model parameters given to the algorithm. In FPWC, the model computed based on the Fourier optics can be used as the initial guess into the algorithm. Since it is pretty close to the true value, the parameter estimation converges within only one or two E-M iterations.}

\subsection{E-M algorithm for FPWC system} \label{sec:derivation}
\textcolor{black}{The state-space model of the FPWC system defines a typical input-output hidden Markov process, where the focal plane electric fields are the hidden variables and the Jacobian matrix as well as the process and measurement noise covariance matrices are the model parameters, so the E-M algorithm is suitable for the this system. Moreover, since the dynamics of different pixels are decoupled in the FPWC system under the linear assumption, we can separately and in parallel identify the model parameters of each pixel separately, which saves a lot of computation time.}

Here we copy the state transition and observation equations defined by Eq.~\ref{eq:observationNoise} and Eq.~\ref{eq:transitionNoise},
\begin{equation} \label{eq:ssm}
\begin{aligned}
x_{k, j} &= x_{k-1, j} + G_j u_k + w_{k, j}, \quad w_{k, j} \sim N(0, Q_{k, j}),\\
z_{k, j} &= H_{k, j} x_{k, j} + n_{k, j}, \quad H_{k, j} = 4 u_k^{pT} G_j^T, \quad n_{k, j} \sim N(0, R_{k, j}).
\end{aligned}
\end{equation}
where $k \in \{1, \cdots, N_{d}\}$ is the index of the control iterations, $j \in \{1, \cdots, N_{pix}\}$ is the index of camera pixels, and $N_d$ is the total number of the control iterations. \textcolor{black}{Based on Eq.~\ref{eq:ssm}, the model parameters, hidden variables and model inputs and observations for the single-pixel E-M algorithms can be respectively denoted as  $\theta_j = \{G_j, Q_{,1:N_d j}, R_{1:N_d, j}\}$, $X_j = \{x_{0:N_d, j}\}$ and $Y_j = \{u_{1:N_d}, u_{1:N_d}^p, z_{1:N_d, j}\}$. By assuming the process noise $w_{k, j} \cong \Delta G_j u_k + r_k$ and the Jacobian errors from different actuators are independent as shown in Eq.~\ref{eq:errstd}, the process noise covariance matrix is
\begin{equation} \label{eq:processNoiseDef}
Q_{k, j} = u_k^T u_k Q_j + S_{k, j} = u_k^T u_k \sigma^2 \mathbb{I}_{2 \times 2} + \delta^2 \mathbb{I}_{2 \times 2},
\end{equation}
where $S_{k, j}$, the covariance from the system instability term $r_k$, is assumed to be a constant scalar matrix, $\delta^2 \mathbb{I}_{2 \times 2}$, over iterations. In our following simulation and experiment, since the instability term is much smaller compared with the Jacobian bias, we neglect $\delta^2$ (assume $\delta^2=0$) and only identify $\sigma^2$ to determine the process noise covariance. Without changing exposure time, the observation noise covariance matrix is also a constant scalar matrix over iterations,
\begin{equation} \label{eq:observationNoiseDef}
R_{k, j} = R_{j} = \nu^2 \mathbb{I}_{n \times n},
\end{equation}
where $\nu$ is the standard deviation of the observation noise and $n$ is the number of pairs of probes. As a result, the model parameters are simplified as $\theta_j = \{G_j, \sigma^2, \nu^2\}$ in the current E-M algorithm for the FPWC system.}

\textcolor{black}{The E-M equations for $X_j$, $Y_j$, and $\theta_j$ can be derived following the approach of Ghahramani et al.\cite{ghahramani1996parameter}. As shown in that paper, for a linear Gaussian dynamical system like FPWC, the E-step can be achieved by Kalman filtering and Rauch smoothing and the M-step can be achieved by finding the analytical solution of a quadratic optimization problem. However, since the Jacobian matrix and observation matrix in FPWC have shared parameters and our control variables are high-dimensional, the model parameter update equations and the optimization method are a little different from the standard approach. The implementation details of the E-M algorithm for FPWC system are explained in next section. For notational simplicity, we will omit the subscript $j$  in the following derivations and discussions, understanding that the E-step and the M-step are repeated $N_{pix}$ times.}

\subsection{Actual implementation} \label{subsec:implementation}
\subsubsection{The E-Step} \label{subsubsec:Estep}
\textcolor{black}{In what follows, we introduce notations $\hat{x}_{k_1|k_2}$ and $P_{k_1|k_2}$, which represent the estimated expectation and covariance of the hidden states at control iteration $k_1$ given observations up to and including at control iteration $k_2$. With these simplified notations, the conditional probability in the E-step becomes
\begin{equation} \label{eq:linearGaussian}
\mathcal{Q}(X)=p(X| Y, \theta)=\prod_{k=1}^{N_d} \mathcal{N}(\hat{x}_{k|N_d}, P_{k|N_d})
\end{equation}
in our linear Gaussian FPWC system. This conditional probability can be derived from a combined approach using Kalman filter and Rauch smoother.}

The Kalman filter first forward propagates the states and estimates the hidden states  based only on the data up to the current step. The Kalman filter optimization problem is defined in Sec.~\ref{subsec:estimationControl}. The solution to the optimization problem gives five Kalman filter equations,
\begin{subequations} \label{eq:kalman}
\begin{align}
&\hat{x}_{k|k-1} = \hat{x}_{k-1|k-1} + G u_k,\\
&P_{k|k-1} = P_{k-1|k-1} + Q_k,\\
&K_k = P_{k|k-1} H_k^T (H_k P_{k|k-1} H_k^T + R_k)^{-1},\\
&\hat{x}_{k|k} = \hat{x}_{k|k-1} + K_k(z_k - H_k \hat{x}_{k|k-1}),\\
&P_{k|k} = (\mathbb{I} - K_k H_k) P_{k|k-1}^{-1},
\end{align}
\end{subequations}
where $\hat{x}_{k|k-1}$ and $P_{k|k-1}$ are the a priori knowledge of the states and covariance matrix from observations up to control iteration $k-1$, and $\hat{x}_{k|k}$ and $P_{k|k}$ are the a posteriori estimates updated by the observations at step $k$.

Rauch smoother then propagates the states backwards from the last step to the starting step and further updates the estimates based on the data of the future steps. Mathematically,  Rauch smoother is a Kalman filter using the next hidden state as the observation. The Rauch smoothing equations are,
\begin{subequations} \label{eq:rauch}
\begin{align}
&L_k = P_{k|k} P_{k+1|k}^{-1},\\
&\hat{x}_{k|N_d} = \hat{x}_{k|k} + L_k (\hat{x}_{k+1|N_d} - \hat{x}_{k+1|k}),\\
&P_{k|N_d} = P_{k|k} + L_k (P_{k+1|N_d} - P_{k+1|k}) L_k^T,
\end{align}
\end{subequations} 
where $\hat{x}_{k|N_d}$ and $P_{k|N_d}$ are the estimated hidden state's expectation and covariance based on all the $N_d$ steps of data, $Y=\{u_{1:N_d}, u_{1:N_d}^p, z_{1:N_d, j}\}$.

\subsubsection{The M-Step} \label{subsubsec:Mstep}
\textcolor{black}{The M-step defines a stochastic maximum likelihood estimation (MLE) problem.} Based on the Markovian structure of Eq.~\ref{eq:ssm}, the log likelihood of the \textcolor{black}{hidden states, model inputs and observations} is,
\begin{equation} \label{eq:likelihoodMstep}
\begin{aligned}
L(G, Q, R) = &\log \ \prod_{k=1}^{N_d} p(z_k|x_k, H_k, R_k) \prod_{k=1}^{N_d} p(x_k|x_{k-1}, u_k, G, Q_k)\\
=&-\frac{1}{2}\sum_{k=1}^{N_d} (z_k - H_k x_k)^T R_k^{-1} (z_k - H_k x_k) - \frac{1}{2} \sum_{k=1}^{N_d}\log \ |2 \pi R_k|\\
& -\frac{1}{2}\sum_{k=1}^{N_d} (x_k - x_{k-1} - G u_k)^T Q_k^{-1} (x_k - x_{k-1} - G u_k) - \frac{1}{2} \sum_{k=1}^{N_d}\log \ |2 \pi Q_k|.
\end{aligned}
\end{equation}
where 
\begin{equation} \label{eq:likelihoodMstep2}
R_k = R,\ Q_k = u_k^T u_k Q,\ H_k = 4 (G u_k^p)^T
\end{equation}

The expectation of this log-likelihood can be calculated  using the state estimates in the E-step. Therefore, we can estimate the model parameters by taking the derivatives of the log-likelihood with respect to each parameter and forcing the resulting expectations to be zero,
\begin{equation} \label{eq:derivative}
\frac{\partial \left<L(G, Q, R)\right>}{\partial G} = 0, \frac{\partial \left<L(G, Q, R)\right>}{\partial Q} = 0, \frac{\partial \left<L(G, Q, R)\right>}{\partial R} = 0.
\end{equation}
This gives the analytical update equations for the model parameters,
\begin{subequations} \label{eq:analytical}
\begin{align}
G = &\{ \sum_{k=1}^{N_d} \frac{1}{u_k^Tu_k}(\hat{x}_{k|N_d} - \hat{x}_{k-1|N_d})u_k^T \notag \\
&+ 4 Q \sum_{k=1}^{N_d}[\hat{x}_{k|N_d}z_k^T - 4(\hat{x}_{k|N_d}\hat{x}_{k|N_d}^T + P_{k|N_d}) G u_k^p]R^{-1}u_k^{pT}\}(\sum_{k=1}^{N_d} \frac{u_k u_k^T}{u_k^Tu_k})^{-1},\\
Q =& \frac{1}{N_d} \sum_{k=1}^{N_d} \frac{1}{u_k^T u_k} [(\hat{x}_{k|N_d} - \hat{x}_{k-1|N_d} - G u_k)(\hat{x}_{k|N_d} - \hat{x}_{k-1|N_d} - G u_k)^T + P_{k|N_d} + P_{k-1|N_d}],\\
R =& \frac{1}{N_d} \sum_{k=1}^{N_d} [(z_k - 4 (G u_k^p)^T \hat{x}_{k|N_d})(z_k - 4 (G u_k^p)^T \hat{x}_{k|N_d})^T + 16 (G u_k^p)^T P_{k|N_d} G u_k^p].
\end{align}
\end{subequations}

Eq.~\ref{eq:analytical} (a) is an implicit equation, so $G$ needs to be found recursively.  From our earlier assumption, $Q$ and $R$ are forced to be scaled identity matrices,
\begin{equation} \label{eq:scalarMatrices}
Q \leftarrow \frac{Tr(Q)}{2} \mathbb{I}_{2 \times 2}, \quad R \leftarrow \frac{Tr(R)}{2} \mathbb{I}_{2 \times 2},
\end{equation}
where we  accordingly obtain
\begin{equation} \label{eq:scalarMatrices2}
\sigma^2 = \frac{Tr(Q)}{2}, \quad \nu^2 = \frac{Tr(R)}{2}.
\end{equation}
The process covariance, $\sigma^2$, can be used in the EFC algorithm for computing the Tikhonov regularization parameter as shown in Eq.~\ref{eq:regularization}.

One shortcoming of this analytical solution is the large matrix inversion in Eq.~\ref{eq:analytical}(a). To ensure the matrix is invertible, we have to collect several thousand steps \textcolor{black}{(greater than the number of actuators on DMs)} of data before making an update, which is unnecessarily time-consuming and also precludes online system adapting. In order to update the model with a smaller amount of data, we can use a stochastic gradient ascent algorithm instead for updating the Jacobian matrix,
\begin{equation} \label{eq:gradientAscent}
G \leftarrow G + \eta \frac{\partial \left<L(G, Q, R)\right>}{\partial G},
\end{equation}
where the tuning parameter $\eta$ defines the learning rate of the algorithm. \textcolor{black}{However, this method may not be able to reach exact optimal solutions.}

These two subsections presented all of the E-M equations for FPWC system. By repeating the iterative E-M approach on all the pixels we can reconstruct the linear state-space model for the entire system. While that is sufficient, it is helpful to apply a final step, forcing the process and observation noise matrices of all the pixels to be equal to their average. Since all pixels in the dark hole share almost the same noise distributions, neglecting the small difference in photon noises, this step enhances the robustness of the E-M algorithm.

\textcolor{black}{The remainder of the paper will present two ways to apply the E-M algorithm to the FPWC system, offline system identification and online adaptive control. In Sec.~\ref{sec:identificationBoth}, we identify the system using precollected data and try to understand the sources of aberrations in our system. In Sec.~\ref{sec:adaptiveControl}, we integrate the E-M algorithm into the control loop, and adapt the model parameters and control policy in real time. Simulation and experimental results are reported in both cases.}

\section{E-M algorithm based system identification} \label{sec:identificationBoth}
\textcolor{black}{In this section, we numerically and experimentally investigate the E-M algorithm based system identification for FPWC. Our goal for the system identification is to precisely characterize the Jacobian errors. In addtion, we will also take this chance to understand important algorithmic details, for example the influence of the hyper-parameters (batch size and amount of data) on the algorithm's performance or how hard it is to characterize different types of model errors.}

\begin{figure}[!htb]
	\centering
	\includegraphics[width=4in]{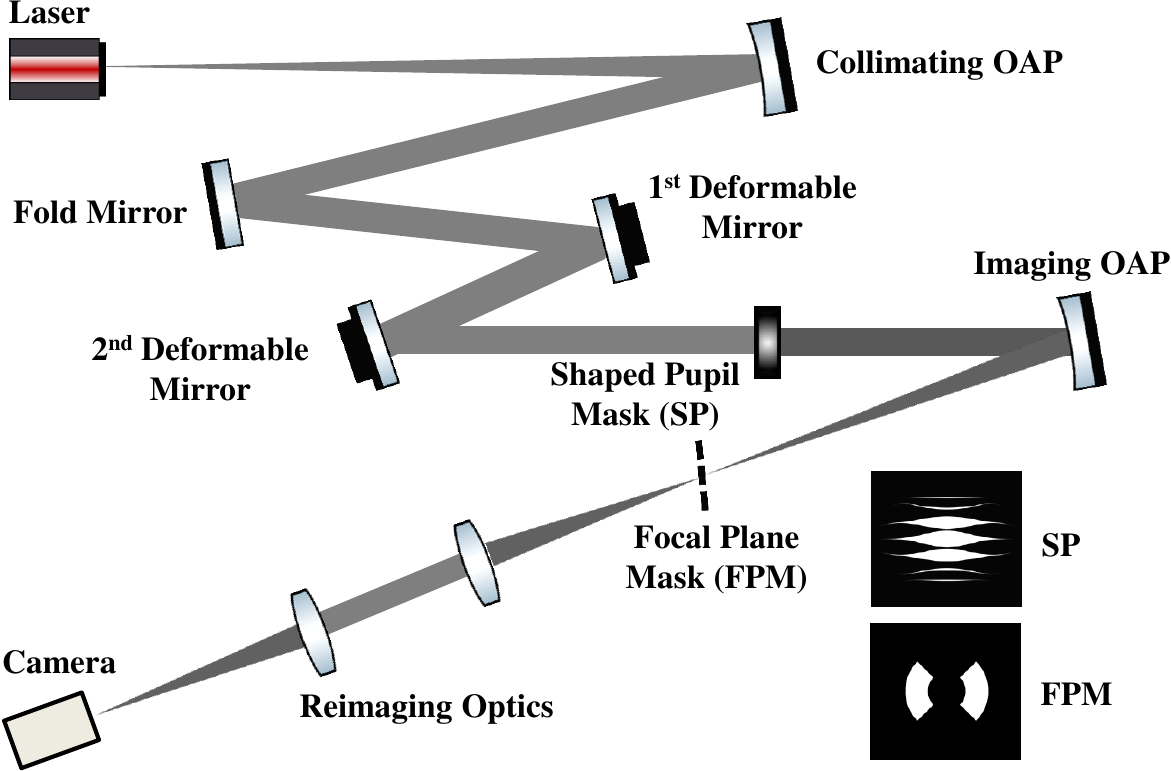}
\caption{Layout of the HCIL testbed. Rippled shaped pupil and bowtie shaped focal plane mask are applied to suppress the contrast in the focal plane. Two Boston MicroMachines MEMS DMs are installed for focal plane wavefront correction.} 
\label{fig:layout}
\end{figure}

\textcolor{black}{The experiment is conducted in the Princeton's High Contrast Imaging Lab (HCIL) and the simulation uses the same setup.} As shown in Fig.~\ref{fig:layout}, the HCIL testbed is a two-DM FPWC system with shaped pupil (SP) coronagraph. It utilizes a ripple pupil plane mask to suppress the contrast by changing the starlight point spread function (PSF). In addition, a bowtie shaped focal plane mask (FPM) blocks the center part of the PSF to avoid camera saturation, which also defines the dark hole regions for the FPWC. Each DM in the HCIL has $952$ actuators. Without loss of generality, we only activate the first DM in simulation and experiment. The second DM is treated as a fold mirror.

\subsection{Numerical verification} \label{sec:simulation}
\subsubsection{Data generation} \label{subsec:dataGeneration}
\textcolor{black}{In the numerical study, we simulated the DM commands and resulting camera images under an imperfect lab condition. Wavefront aberrations with 10nm RMS were added to the shaped pupil plane and two DM planes. The DM actuators' gains were biased by $20\%$ to account for the influence function errors.\footnote{The influence function shape errors were neglected in our simulation, however, the E-M algorithm is able to handle this type of errors as proved in the experimental results.} Shot noises and readout noises were added to the simulated camera images, where the noises' standard deviations were chosen based on  measurements in the HCIL. In the numerical model, the masks are modeled as 0-1 binary matrices, the propagations through the OAPs or lenses to their focuses are modeled as Fourier transform, and all the free space propagations between devices are modeled as Fresnel propagations.}

To sufficiently explore the controllable space of the DM, we generated the data by applying random DM commands in the system identification approach. In our simulation, in total $4000$ random voltage commands (between $-0.6 - 0.6$ volts) were applied to the DM and the resulting camera images were simulated. A fixed exposure time of $0.1$ sec was used for the camera images. For each random DM commands, we collected two pairs of probing images, so we have in total 16000 images ($2 \text{ images/pair} \times 2 \text{ pairs/command} \times 4000 \text{ commands}$) in our data set. The random commands between $-0.6$ and  $0.6$ volts typically result in contrast changes at a level of $1 \times 10^{-6}$. In order to make the DM influence significant enough for learning so that the effect is larger than the  background speckles, in our simulation, we first ran wavefront control for four steps to reach a contrast of roughly $3 \times 10^{-6}$ and then applied the random DM commands and generated the images. Same ``probe" comands were used for all $4000$ data points. Although identical pair-wise probes are not necessary for the E-M system identification, as as will be discussed, it helps us build a metric to evaluate the effectiveness of the identification.

\subsubsection{Evaluation metrics of the identification accuracy} \label{subsec:metrics}
Three metrics were used to evaluate the model errors in our analysis. \textcolor{black}{The first is the percentage error of the E-M identified Jacobian, $G_{EM}$, compared with the true Jacobian including optical aberrations and influence function biases, $G$,
\begin{equation} \label{eq:JacobianError}
\text{Jacobian Error} = \frac{\| G_{EM} - G\|_2^2}{\|G\|_2^2} = \frac{\| \Delta G_{EM}\|_2^2}{\|G\|_2^2}.
\end{equation}}

The second metric assumes we are blind to the true Jacobian matrix (which is true in the experiment); we thus reserve part of the data as a validation set. Theoretically, the difference between two neighboring observations with the same probing commands is a function of only the DM commands,
\begin{equation} \label{eq:validationErr}
\Delta z_k = z_k - z_{k-1} = 4 (G u^p)^T (x_k -x_{k-1}) = 4 (Gu^p)^T G u_k,
\end{equation}
\textcolor{black}{so we can define a percentage validation error of the identified Jacobian matrix, via
\begin{equation} \label{eq:validationErr2}
\begin{aligned}
\text{Validation Error} &= \frac{\sum_{k = 1}^{N_{v}} \|\Delta z_k - 4 (G_{EM} u^p)^T G_{EM} u_k\|_2^2}{\sum_{k=1}^{N_v} \|\Delta z_k\|_2^2}\\
& =  \frac{\sum_{k = 1}^{N_{v}} 4 \|u^{pT} (G^T G - G_{EM}^T G_{EM}) u_k\|_2^2}{\sum_{k = 1}^{N_{v}} 4 \|u^{pT} G^T G u_k\|_2^2},
\end{aligned}
\end{equation}
where $N_v$ is the number of data steps in the validation set. The scale of validation error could be a little different from Jacobian error since it actually measures the difference between $G_{EM}^T G_{EM}$ and $G^T G$ instead of $G_{EM}$ and $G$, however, they should have similar trends and are both good indicators of model accuracy.}

\textcolor{black}{The third metric that indirectly reflects the accuracy of a Jacobian matrix is the correction speed and the final achievable contrast of the wavefront control using it. With a more accurate Jacobian matrix, the wavefront control should achieve a higher contrast with fewer control iterations.}

\subsubsection{System identification results} \label{subsec:IDresults}
\textcolor{black}{In this section, we applied the E-M algorithm based system identification in various ways to the simulation data to test the algorithm. The analytical method in Eq.~\ref{eq:analytical} and the gradient ascent method in Eq.~\ref{eq:gradientAscent} were repectively tried to solve the stochastic MLE problem. For the gradient method, we also examined the effect of using different batch sizes. The batch size is a machine learning term referring to the number of data points utilized in one E-M update. Theoretically, small batch sizes enable timely model parameter updates and time-efficient parallel computing, but sacrifice the accuracy of each update because the hidden states estimation with small batch sizes has relatively larger covariance. The algorithm was also investigated with different numbers of data points. Our goal for this section is mainly to validate the reasonability of the evaluation metrics defined in the previous section, and to compare the performance of the algorithm given different optimization methods, batch sizes and amount of data using these metrics.}

\begin{figure}[!htb]
	\centering
	\includegraphics[width=6.3in]{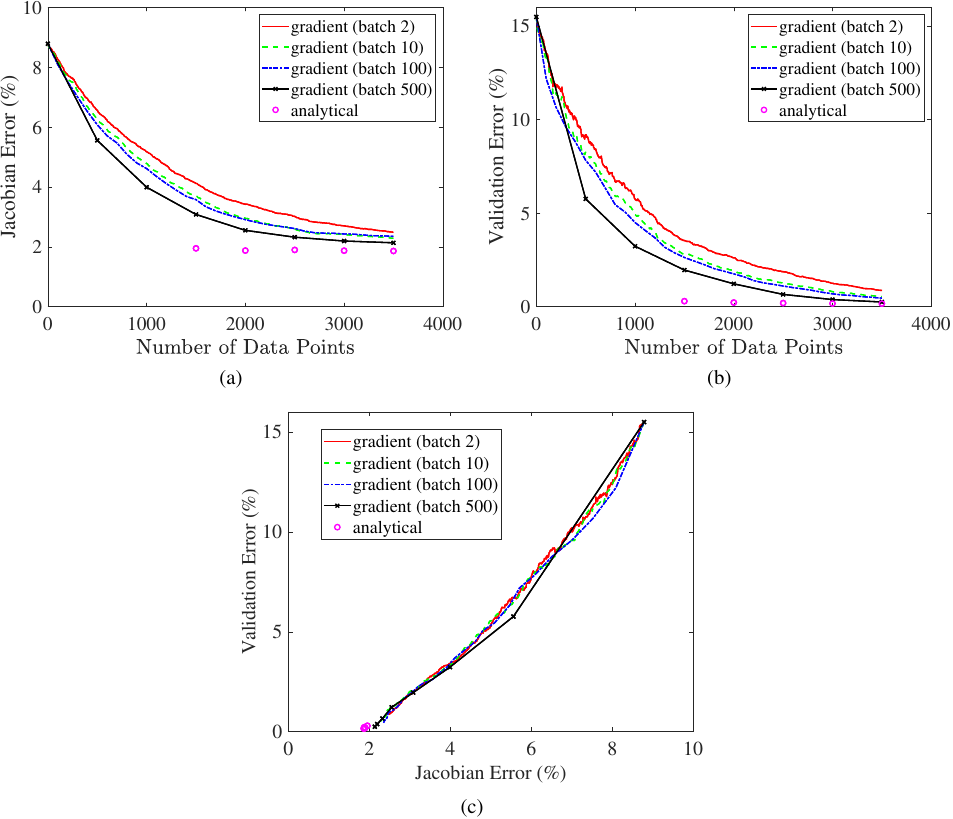}
\caption{(a) Jacobian errors, (b) validation errors and (c) their relations from a simulation over the number of data points in the training set. Different methods, including analytical solutions and stochastic gradient ascent solutions with the batch sizes of 2, 10, 100, and 500 data points are compared using the simulated training data.}
\label{fig:identificationSim}
\end{figure}

Figure~\ref{fig:identificationSim} shows the change in the Jacobian errors and the validation errors with respect to the number of data points. We saved the last $500$ steps of data for validation, so at most $3500$ data points were used for system identification. Results using the analytical method and the gradient ascent method with the batch sizes of 2, 10, 100, 500 are reported. As shown in the figure, the validation error curves resemble the Jacobian error curves, validating it a good metric of model accuracy in the experiment. The stochastic descent algorithm works with a wide range of batch sizes all with similar validation errors, though too small a batch size underperforms compared with others. The analytical method does not work with fewer than $1500$ data points because of the ill-posed matrix inversion in Eq.~\ref{eq:analytical} (a). However, it outperforms the gradient ascent once given enough data. The identification accuracy primarily depends on the number of data points used, no matter what optimization methods or batch sizes we apply.

\begin{figure}[!htb]
	\centering
	\includegraphics[width=3.06in]{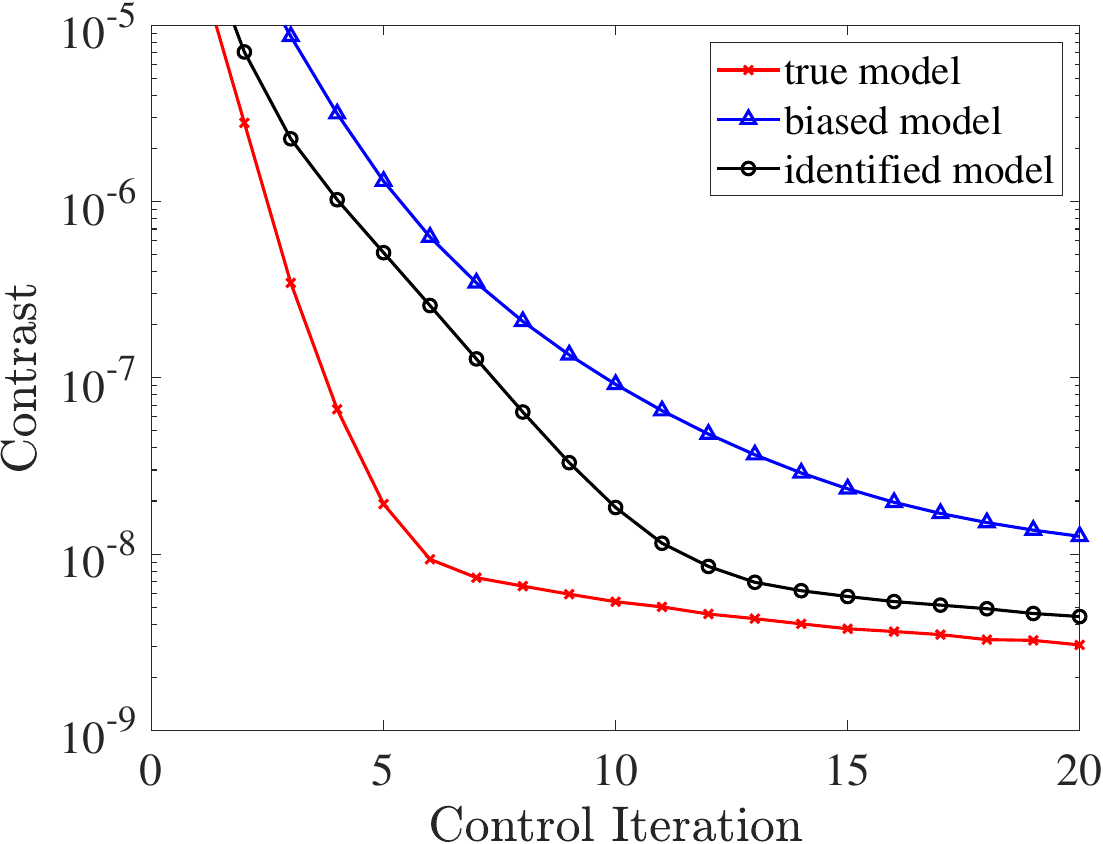}
\caption{Contrast curves of simulated wavefront correction in HCIL. Biased physics model, true model and identified model using analytical method ($3500$ data points) are tested respectively.} 
\label{fig:identificationControlSim}
\end{figure} 

Figure~\ref{fig:identificationControlSim} shows the simulated wavefront correction using the original biased model \textcolor{black}{(computed using Fourier optics with no knowledge of the true aberrations)}, the true model \textcolor{black}{(computed using Fourier optics with full knowledge of the true aberrations)} and the best identified model (analytical solution using $3500$ data points). EFC with a fixed regularization parameter and batch process estimation with two pairs of probing commands were used in this simulation. As can be seen, the identified model beats the biased model in both the wavefront control speed and the final contrast. The contrast gap between the true model and the biased model is significantly reduced after the E-M system identification.

\subsection{Experimental results} \label{sec:identification}
\subsubsection{Data collection} \label{subsec:dataCollection}
The same sampling policy was used in experiment as in simulation: we ran the wavefront correction to reach a relatively high contrast (settling at around $3 \times 10^{-6}$), applied $4000$ random DM commands (between $-0.6$ and $0.6$ volts) and collected the resulting difference images, saving the last $500$ steps as the validation set. Again, two pairs of DM probes were used for observation at each step.

\subsubsection{Identification results} \label{subsec:IDandControl}
\begin{figure}[!htb]
	\centering
	\includegraphics[width=3.06in]{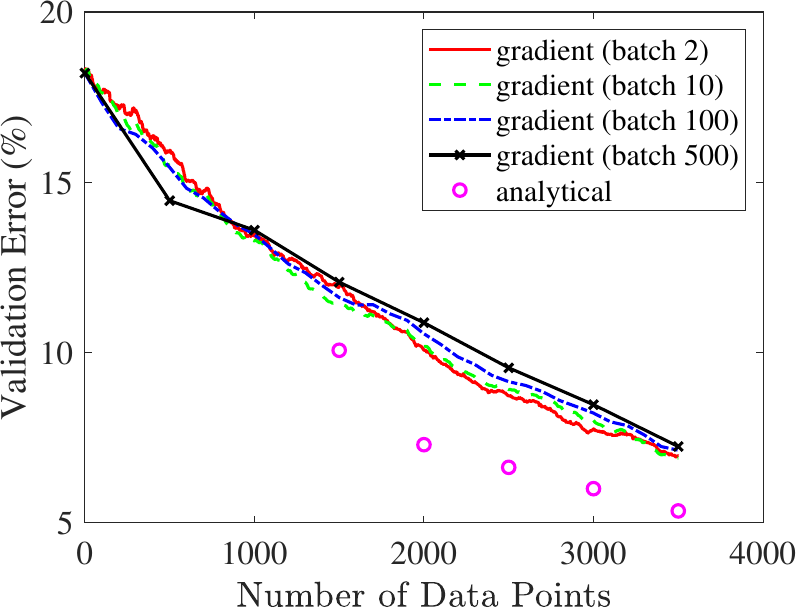}
\caption{Validation errors in the experiment over the number of data steps in the training set. Different methods, including analytical solutions and stochastic gradient ascent solutions with the batch sizes of 2, 10, 100, 500, are compared using the experimental data.} 
\label{fig:validationErr}
\end{figure}

\textcolor{black}{With the validation error proved to be a good metric, now we use this metric to evaluate the identifcation results with the experimental data. As shown in Fig.~\ref{fig:validationErr}, the validation error curves of various cases decrease with the same trends as in Fig.~\ref{fig:identificationSim} (b), showing that the E-M algorithm also successfully detects and corrects the Jacobian errors in the experiment. }

\textcolor{black}{Further analysis of the sources of Jacobian errors in experiment can be found in Appendix~\ref{subsec:aberrationSource}. As shown by this regression analysis of the identified Jacobian, DM actuator's gain errors and pupil plane wavefront phase aberrations explain around half of the model errors in our experiment.\footnote{\textcolor{black}{Other errors may be the influence function shape errors, the wavefront aberrations on the plane of other devices and the system nonliearities beyond the algorithm's identfication ablity.}} Among thees factors, the DM gain errors are easily corrected with only a few of data, while the wavefront aberrations are corrected slower and also varies over time.}

\begin{figure}[!htb]
	\centering
	\includegraphics[width=6.3in]{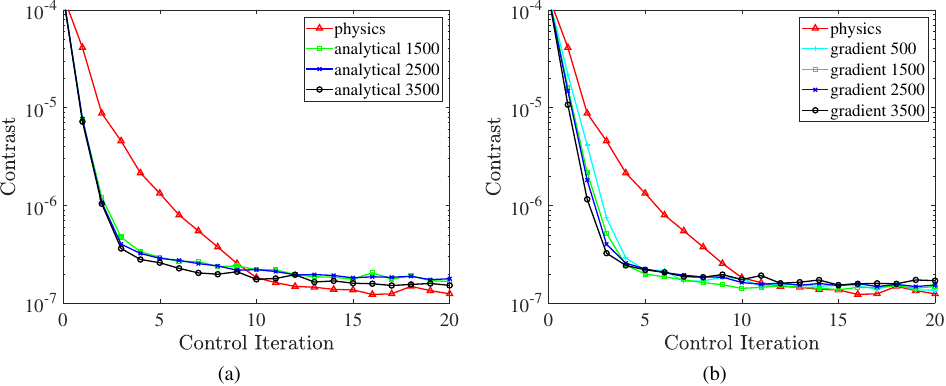}
\caption{Measured contrast in the HCIL over the control iterations. (a) wavefront corrections using physics model and analytical identified Jacobians with 1500, 2500 and 3500 data points. (b) wavefront corrections using physics model and gradient ascent identified Jacobians (bath size of 500) with 500, 1500, 2500 and 3500 data points.} 
\label{fig:control}
\end{figure}

\textcolor{black}{We also compared the wavefront control results using the identified model and the original/biased physics model. In the physics model, we had no knowledge of the wavefront aberrations and assumed the same gain and influence function shape for all the actuators. Similarly, EFC and batch process estimatiton were used in all the wavefront correction trials. Figure~\ref{fig:control} (a) and (b), respectively, show the wavefront control curves (contrast vs. control iteration) using the analytical Jacobian solutions and the gradient ascent Jacobian solutions with different amount of data.} In both cases, the wavefront corrections with the identified models are much faster than the biased physics model in the early stage; they all reached a contrast better than $3 \times 10^{-7}$ within only four to five control iterations. However, the analytical Jacobians did not perform better than the gradient ascent solutions as expected. After reaching a high contrast, the analytical Jacobians experienced some difficulties in correcting the small residual aberrations, resulting in a final contrast slightly worse than the physics model. We speculate that the analytical E-M solutions are overfitted to the data noise. In contrast, the gradient ascent solutions reached the same ultimate contrast as the physics model. This is mainly because the achievable final contrast in the lab is currently limited by the scattered, incoherent light. On conclusion from these results is that the gradient method is better for experimental applications. In addition, the wavefront correction speed did not improve much as the number of data points increased. \textcolor{black}{This may be because the key factors that influence the wavefront correction speed, probably the DM actuator's gain errors as discussed in the appendix, were detected and corrected with only tens of data points and/or offline system identification didn't handle the time-varying data well.}

\section{E-M algorithm based adaptive control} \label{sec:adaptiveControl}

The experimental results in Sec.~\ref{sec:identificationBoth} demonstrated the ability of the E-M algorithm to improve the Jacobian accuracy, even with only small amount of data. However, this system identification workflow (data collection - identification - wavefront correction) cannot keep up with some of the most important time-varying errors, such as thermally induced phase aberrations. In this section, we present an E-M algorithm based real-time adaptive control framework, or more specifically a reinforcement learning control framework, to solve this problem. \textcolor{black}{This reinforcement learning control strategy is not fundamentally different from the E-M algorithm based system identification; we use the same algorithm developed in Sec.~\ref{sec:EM} but only directly feed the wavefront correction data instead of the precollected data with random DM commands into the E-M equation.}

\subsection{Reinforcement learning for FPWC} \label{subsec:reinforcementLearning}

Reinforcement learning control has attracted much attention recently as an important branch of  machine learning. In reinforcement learning, the system, or agent, alternately runs a control policy to explore the environment and an adaptation step that varies the policy based on the information from the control step. Since the agents directly learn from the control attempts, it is more efficient for them to find the best control policies and track the model variations in real time. This technique has been widely applied to training complex control systems, such as those playing the game of Go \cite{silver2016mastering} or video games\cite{mnih2013playing},  robot manipulation, motion planning, and locomotion\cite{kober2013reinforcement}. 

\begin{figure}[!htb]
	\centering
	\includegraphics[width=6in]{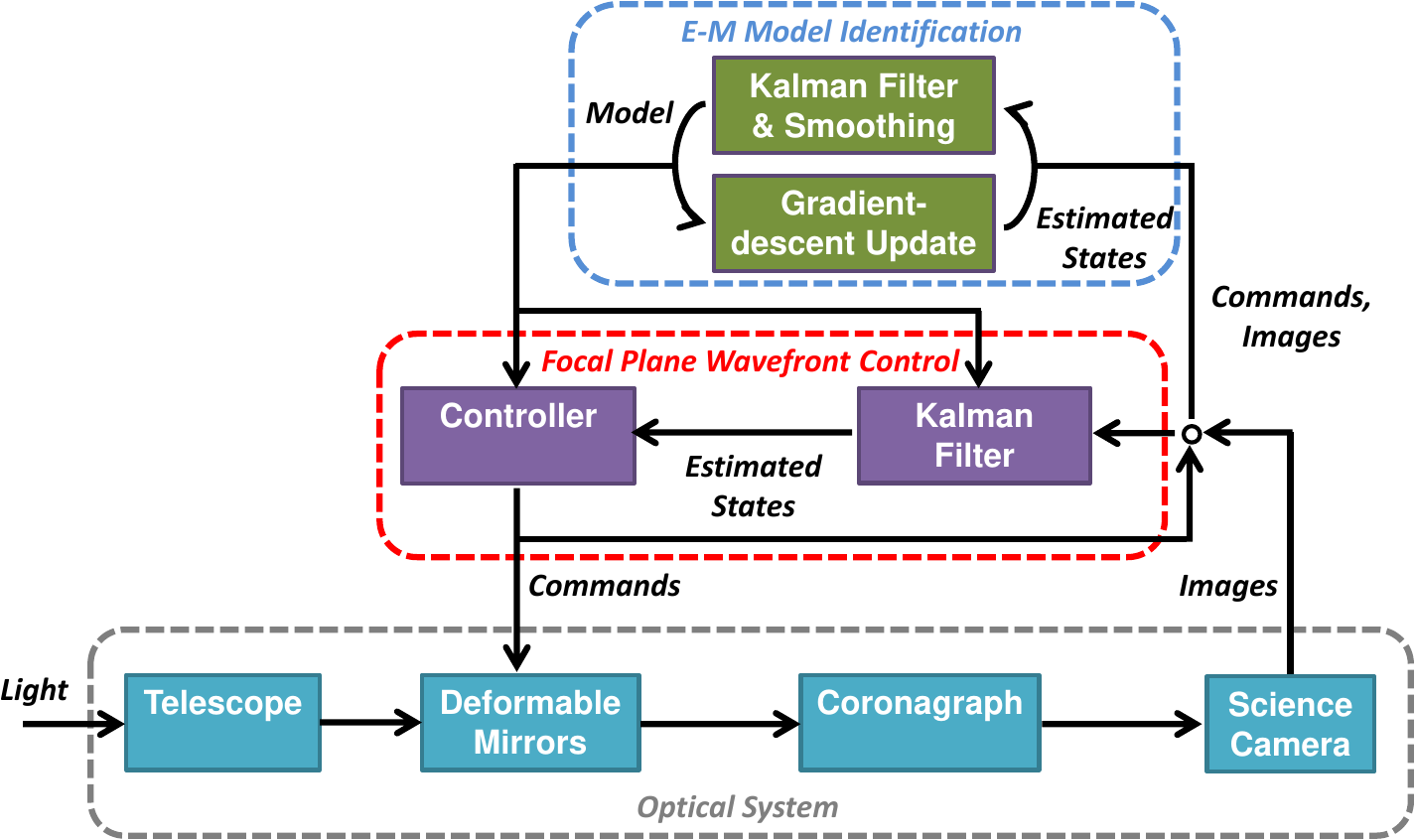}
\caption{Block diagram of the E-M algorithm based adaptive focal plane wavefront correction (FPWC) system.} 
\label{fig:adaptiveControl}
\end{figure}

Figure~\ref{fig:adaptiveControl} shows the block diagram of the proposed adaptive FPWC system. It combines the wavefront estimation and control with the E-M system identification presented in Sec.~\ref{sec:EM}. In this scheme, we no longer use random DM commands for identification. Instead, the DM commands and resulting images from the control loops are sent to the E-M algorithm to update the model  instantanesouly. The new adaptive FPWC system now loops between running steps of wavefront estimation and control and updating the model parameters (which also means updating the control and estimation policy). In addition, not only is the Jacobian matrix, $G$, identified in the adaptive/reinforcement learning control step, so too are the  process noise, $\sigma^2$, and observation noise, $\nu^2$ as demonstrated in Eq.~\ref{eq:scalarMatrices} and Eq.~\ref{eq:scalarMatrices2}. These are then used to tune the wavefront estimator (the covariance matrices of process noises and observation noises in Kalman filter) and controller (Tikhonov regularization matrix in EFC) based on Eq.~\ref{eq:processNoiseDef}, Eq.~\ref{eq:observationNoiseDef} and Eq.~\ref{eq:regularization}.\footnote{In our software implementation, we introduce a hyperparameter, $\gamma$, to Eq.~\ref{eq:regularization}, which defines a modified regularization matrix, $W^{\prime} = \gamma W = 2 \gamma N_{pix} \sigma^2 \mathbb{I}$, because we found the controller is usually able to be more aggressive than the theoretical suggestion.} As a result, the Kalman filter estimator  better balances the weights of the model predictions and observations, and the controller better chooses the damping parameter in the wavefront correction.

\subsection{Reinforcement learning simulation} \label{subsec:adaptiveSim}
\begin{figure}[!htb]
	\centering
	\includegraphics[width=6.3in]{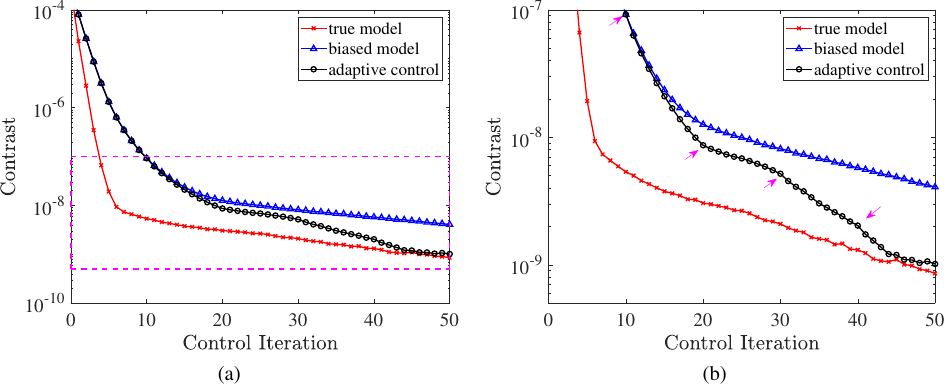}
\caption{(a) Simulated FPWC reinforcement learning control and the benchmark wavefront control with the true Jacobian matrix and the fixed biased Jacobian matrix. The model parameters are updated using the E-M algorithm every ten iterations. (b) Zoomed-in figure of the box region in (a). The E-M identifications occurred at the iterations marked by pointed arrows.} 
\label{fig:adaptiveSim}
\end{figure}

Again using the imperfect lab conditions that result in phase aberrations and actuator gain biases as stated in Sec.~\ref{subsec:dataGeneration}, we simulated the reinforcement learning control for $50$ control iterations. Two pairs of probing images were collected at each iteration for wavefront estimation. In Sec.~\ref{sec:identificationBoth}, we used same pair-wise probes for the convenience of validation error calculation, however, here we allowed the DM probes to vary among different control iterations in the reinforcement learning control simulation. After every $10$ control iterations, we supplied the control commands ($10$ steps), the pair-wise probes ($2$ pairs/step $\times 10$ steps) and  the camera images  ($2$ images/pair $\times 2$ pairs/step $\times 10$ steps) to the E-M algorithm to update the Jacobian matrix and the tuning parameters in the estimator and controller. For comparison, the wavefront control with the true Jacobian model and the fixed biased Jacobian model were also simulated. In both of these benchmark cases, the Kalman filter and the EFC controller were tuned to the best manually.  Figure~\ref{fig:adaptiveSim} shows the results of the three simulations.  As can be seen, the reinforcement learning control gradually closed the contrast gap between the biased model and the true model. The E-M adaptation at every ten iterations can be clearly seen on the correction curves.

\subsection{Reinforcement learning experiment in HCIL} \label{subsec:adaptiveLab}

In this section we present the results of using the reinforcement learning adaptive control approach in the HCIL.  Unfortunately, because the ultimate contrast achievable in the HCIL is limited to roughly $1.5 \times 10^{-7}$ due to incoherent  background light (as seen in Figs.~\ref{fig:control}), it is not possible to reproduce the simulation results from the previous section.  There, the adaptation step was run after each 10 iterations of the control.  But as can be seen in Fig.~\ref{fig:adaptiveSim}, the modeled system reaches a contrast better than the lab limit of $10^{-7}$ in fewer than 10 steps, before the first reinforcement learning step.  Through trial and error it was found that the E-M algorithm cannot robustly identify the system with fewer than 10 learning steps.  Therefore, to experimentally verify the algorithm, we limited each FPWC run to 10 control iterations and updated the model parameters using the E-M algorithm after each trial. The  Jacobian and tuning parameters were then used for the next trial of wavefront correction.

\begin{figure}[!htb]
	\centering
	\includegraphics[width=6.3in]{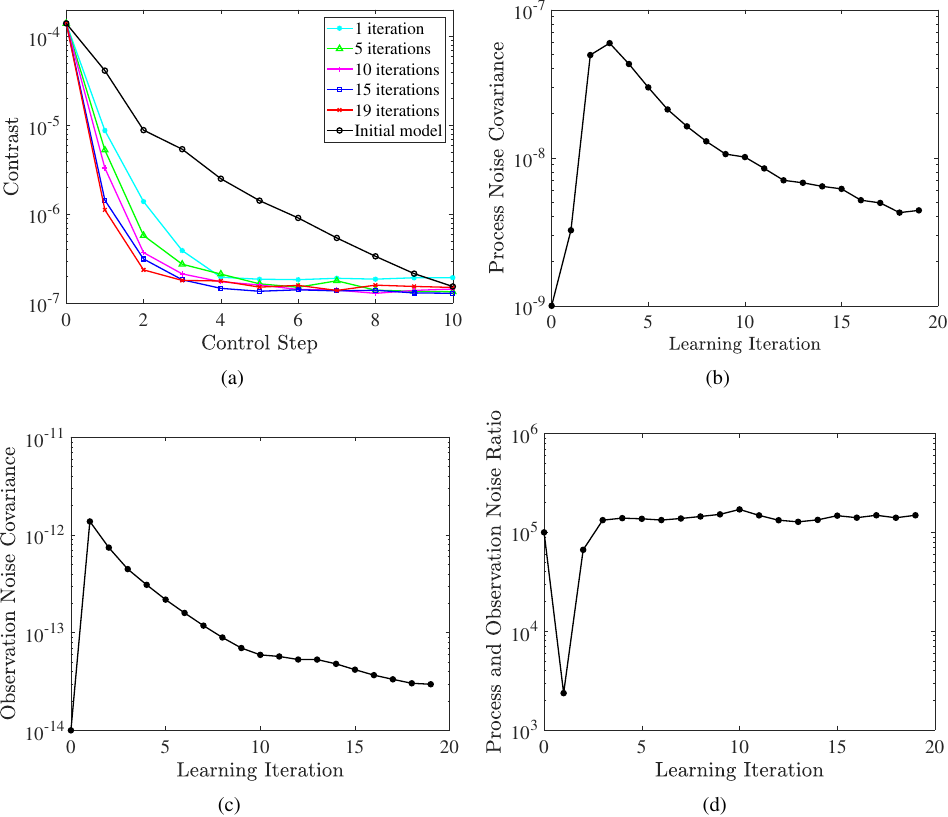}
\caption{Change of (a) the wavefront correction speed, (b) the process noise, (c) the observation noise and (d) the process and observation noise ratio with respect to the learning iterations. To compare the wavefront correction speed, we present the measured contrast over 10 control iterations for the initial model and identified model after 1, 5, 10, 15, 19 learning iterations.}
\label{fig:learning}
\end{figure}

As shown in Fig.~\ref{fig:learning} (a), the rate of convergence of the wavefront correction became faster after each learning trial. Note that we ran the E-M identification after every learning iteration, however, to keep the figure clean, we only report a few of the typical results (wavefront control with the initial biased model and after 1, 5, 10, 15, 19 learning trials). After only 19 learning trials, the FPWC system was able to reach $1 \times 10^{-6}$ in one control step and below $2 \times 10^{-7}$ contrast in three control steps, which is faster than the results from the off-line system identification. \textcolor{black}{This indicates the wavefront control provided more informative data compared with random DM commands. One possible explanantion is that the controller in wavefront correction more frequently moves the DM actuators not blocked by the coronagraph masks, and the parameters of these actuators (corresponding columns of the Jacobian matrix) are actually the key parts to improve the wavefront correction. As a contrast, the random command policy indistinguishably moves all the actuators, which may not be efficient. The reinforcement learning framework may also have captured some time-varying errors. However, since our testbed is pretty stable over short time intervals, this should not be the main reason that the reinforcement learning control outperformed the system identification.}

\begin{figure}[!htb]
	\centering
	\includegraphics[width=4.5in]{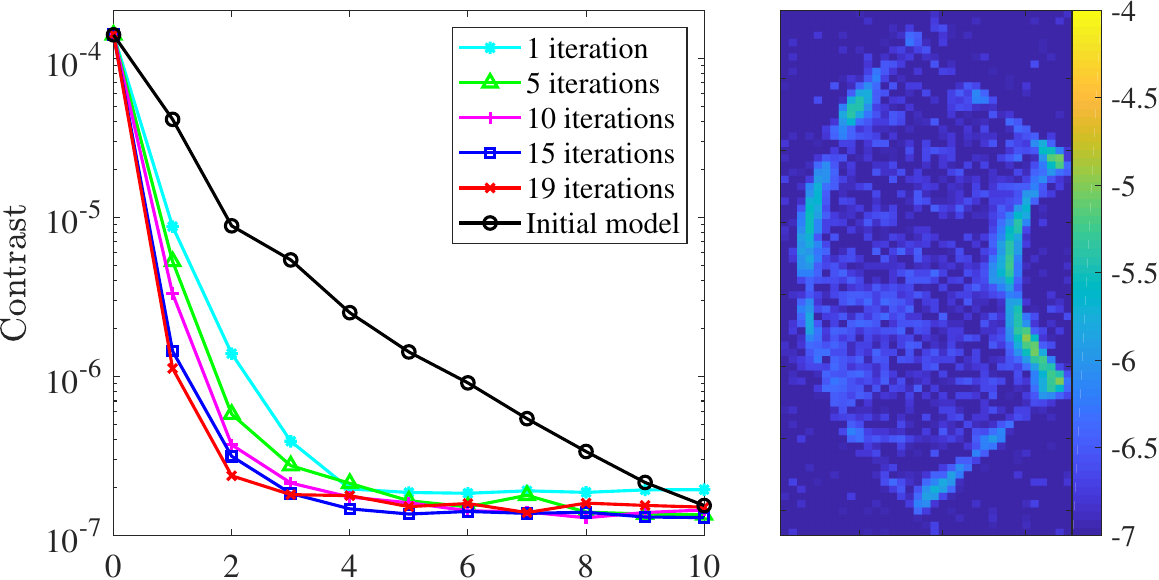}
\caption{A still image from the video about the adaptive wavefront correction in HCIL. (MP4, 1.34 MB)} 
\label{fig:video}
\end{figure}

Figs.~\ref{fig:learning} (b) (c) (d) show the changes in the estimates of process noise and observation noise covariances and their ratio at each  learning trial. As shown in these figures, we underestimated the noise levels at the beginning. The adaptive controller quickly corrected these incorrect assumptions. Then, the adaptive controller gradually corrected the errors in the Jacobian matrix, so that the process and observation noise covariance estimates decreased with additional learning trials. More details about the adaptive control experiment can be seen in the video in Fig.~\ref{fig:video}.

By using this reinforcement learning approach, much effort is saved, and accuracy gained, by not having to take testbed layout measurements, perform phase retrieval and surface characterization, or having to manually tune the controller and estimator parameters. The reinforcement learning adaptive control results also shows promise for enabling self-maintenance of the FPWC during the mission.

\section{Conclusion and future work} \label{sec:conclusion}
Efficient and successful focal plane wavefront control and estimation in coronagraph instruments requires  accurate modeling of the optical system. In this work, we first proposed an expectation-maximization (E-M) algorithm to identify the optical system as a linear state-space model. According to the simulation and experimental results in the Princeton HCIL, the algorithm successfully corrects model errors such as those produced from errors in the DM gains and initial phase aberrations.  Use of the identified models significantly increases the rate at which the wavefront correction converges. We also developed a model based adaptive/reinforcement learning control scheme based on this E-M algorithm. The adaptive controller alternates between the wavefront correction and the model parameter self-adaptation, which significantly improves the performance of both the estimator and controller and requires only tens of learning iterations. This approach is very promising for the automatic maintenance of the FPWC system in future space missions.

Future work will focus on generalizing this frame work with more realistic assumptions. First, we plan to identify the full matrix regularization suggested in Sec.~\ref{sec:stochasticOpt} instead of the scalar regularization. This will help us understand the inter-actuator couplings that are neglected by electric field conjugation and stroke minimization, as well as improve the performance of the wavefront correction. Second, we also plan to drop the linearity assumption, and \textcolor{black}{use EKF and neural networks to approximate the optical system as a nonlinear system. The linear assumption does not hold when we need large DM surface chages to correct the influences from telesocpe struts and/or segmented apertures.} By introducing system nonlinearities back into the model, we should be able to further increase the speed and efficiency of the wavefront corrections, gain a deeper contrast, and better extract the exoplanet signal.

\appendix

\section{Regression analysis of the sources of Jacobian Errors} \label{subsec:aberrationSource}
The Fourier optics analysis in Eq.~\ref{eq:stateTransition} shows that the Jacobian errors primarily come from errors in the pupil field, $E_{ab}$, and the influence functions, $f_{1:N_{act}}$. Thus, we can analyze the sources of the Jacobian errors by fitting $E_{ab}$ and $f_{1:N_{act}}$ to our identified Jacobian matrix, $G_{EM}$. After rearranging the real-valued Jacobian matrix, $G_{EM}$, back into the complex form, $F_{EM}$, based on Eq.~\ref{eq:realState}, the fitting problem can be formulated as,
\begin{equation} \label{eq:fit}
\min_{E_{ab}, f_{1:N_{act}}} \quad \| F(E_{ab}, f_{1:N_{act}}) - F_{EM} \|_F^2.
\end{equation}
The pupil electric field and influence functions are respectively parameterized as,
\begin{equation} \label{eq:parametrize}
\begin{aligned}
E_{ab} &= \exp(i \sum \beta_m Z_m) \approx 1 + i \sum \beta_m Z_m, \\
f_{q} &= \rho_q f, \ \forall q = 1, \cdots, N_{act},
\end{aligned}
\end{equation}
where $Z_m$ and $\beta_m$ are the Zernike polynomials and their coefficients, $f$ is the shape of the influence function, and $\rho_q$ are the actuator gains. For simplicity, this parameterization neglects  amplitude wavefront aberrations and the difference of influence function shapes among actuators. With this parameterization and Taylor expansion in Eq.~\ref{eq:parametrize}, the fitting problem in Eq.\ref{eq:fit} becomes a simple linear, least-square regression in the parameters $\beta_m$ and $p_q$.  

\begin{figure}[!htb]
	\centering
	\includegraphics[width=6.3in]{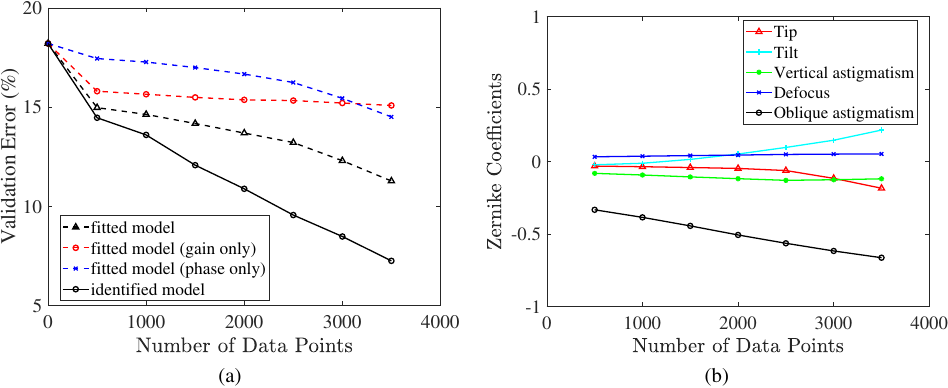}
\caption{Regression analysis of the experimental data in Sec.~\ref{sec:identification}. (a) Validation errors of the E-M identified models and the corresponding fitted models. Validation errors of the fitted models that correct only DM gain errors (red), only phase aberrations (blue) are also reported. (b) First five fitted Zernike coefficients from the regression.} 
\label{fig:fitting}
\end{figure}

\textcolor{black}{Figure~\ref{fig:fitting}(a) compares the validation errors of an identified model (gradient ascent solution with batch size of 500 in Sec.~\ref{sec:identification}) and its fitted model. The validation errors from only fitting with the DM gains or Zernike phase aberrations are also reported. As shown, the fitted model explains more than half the model errors identified by the E-M algorithm, which in part proves our guess about the major sources of model errors. More interestingly, the DM gains are accurately characterized with only the first 500 data points, so the corresponding validation error curve (red) decreases rapidly at the beginning, but changes little as the amount of data increases. In contrast, the validation error from the phase aberrations regression (blue) keeps decreasing as the amount of data increases without reaching plateu. This indicates that the phase aberrations are hard to to correct and may be slowly changing while collecting the data, so the identification algorithm keeps adjusting the Zernike coefficients as the data amount increases. Actually, the curve slope becomes even sharper in the end, because the data in the end may have more similar pupil aberrations as the validation data. The first five fitted Zernike coefficients with respect to the number of data points are further reported in Fig.~\ref{fig:fitting}(b). The defocus and vertical astigmatism do not change much, while the tip, tilt and oblique astigmatism vary over time, which satisfies our observation that the center of the PSF shifted for one pixel horizontally and vertically respectively in our experiment after collecting 4000 data points. This explains why the marginal benefit of data decreases. Moreover, this also justifies the advantage of adapting the system in real time.}

\acknowledgments 
This work was performed under contract to the Jet Propulsion Laboratory of the California Institute of Technology, award number AWD1004079, and under contract to NASA Goddard Space Flight Center, award number AWD1004730.


\bibliography{report}   
\bibliographystyle{spiejour}   


\vspace{2ex}\noindent\textcolor{black}{He Sun} 
is a PhD candidate of Mechanical and Aerospace Engineering at Princeton University. He received his B.S. degree in Engineering Mechanics and Economics from Peking University in 2014. His research interests include coronagraph design and adaptive optics for exoplanet imaging, optimal control and estimation, statistical learning, and robotics. He is a member of the American Astronomical Society and the SPIE.

\vspace{2ex}\noindent\textcolor{black}{N. Jeremy Kasdin}
is a Professor of Mechanical and Aerospace Engineering at Princeton University. He is the Principal Investigator of Princeton's High Contrast Imaging Laboratory and Coronagraph Adjutant Scientist for WFIRST, the Wide Field InfraRed Survey Telescope. He received his Ph.D. from Stanford University in 1991. Professor Kasdin's research interests include space systems design, space optics and exoplanet imaging, orbital mechanics, guidance and control of space vehicles, optimal estimation, and stochastic process modeling. He is an Associate Fellow of the American Institute of Aeronautics and Astronautics and member of the American Astronomical Society and the SPIE.

\vspace{2ex}\noindent\textcolor{black}{Robert Vanderbei}
is a Professor of Operations Research and Financial Engineering at Princeton University. He also holds courtesy appointments in the Department of Mathematics, Astrophysics, Computer Science, and Mechanical and Aerospace Engineering. He received his Ph.D. from Cornell University in 1981. He is a Fellow of the American Mathematical Society (AMS), the Society for Applied and Industrial Mathematics (SIAM) and the Institute for Operations Research and the Management Sciences (INFORMS).

\listoffigures
\listoftables

\end{spacing}
\end{document}